\documentclass[aps,pre,noshowpacs,onecolumn]{revtex4}

\usepackage{pstricks,pst-node}
\usepackage{amscd}
\usepackage{psfrag}
\usepackage{graphicx}
\usepackage{dcolumn}
\usepackage{bm}

\begin{document}

\title{Statistical Mechanics of Unbound Two Dimensional Self-Gravitating Systems}

\author{Tarc\'{i}sio N. Teles}

\email{teles@if.ufrgs.br}

\author{Yan Levin}

\email{levin@if.ufrgs.br}

\author{Renato Pakter}

\email{pakter@if.ufrgs.br}

\author{Felipe B. Rizzato}

\email{rizzato@if.ufrgs.br}

\affiliation{Instituto de F\'{i}sica, UFRGS, Caixa Postal 15051, CEP 91501-970 Porto Alegre, Rio Grande do Sul, 
Brazil}

\begin{abstract}
We study, using both theory and molecular dynamics simulations,
the relaxation dynamics of a microcanonical two dimensional self-gravitating system. 
After a sufficiently large time, 
a gravitational cluster of $N$ particles relaxes to the Maxwell-Boltzmann distribution. 
The time to reach the thermodynamic equilibrium, however, scales with the number of particles.    
In the  thermodynamic limit, $N\rightarrow\infty$ at fixed total mass, equilibrium state is never
reached and the system becomes trapped in a non-ergodic stationary state.   An analytical
theory is  presented which allows us to quantitatively described this final stationary state, 
without any adjustable parameters.
\end{abstract}

\maketitle

\section{Introduction}

Systems interacting through long-range forces behave very differently from those in which particles interact
through short-range potentials. 
For systems with short-range forces, for arbitrary initial condition,  
the final stationary state corresponds to the thermodynamic equilibrium
and can be  described equivalently by either microcanonical, canonical, or grand-canonical ensembles.   
On the other hand, for systems with unscreened long-range interactions, equivalence between 
ensembles breaks down~\cite{Gibbs,Barre2001}.
Often these systems are  characterized by a
negative specific heat~\cite{Ly67,Thir70,Ly77} in the microcanonical ensemble and a broken 
ergodicity~\cite{Muka2005,Rami2008}.  In the infinite particle limit, $N \rightarrow \infty$, 
these systems never reach the thermodynamic equilibrium and become trapped 
in a stationary out of equilibrium state (SS)~\cite{Kl54,Bouch2005}.  Unlike normal thermodynamic equilibrium, the SS
does not have Maxwell-Boltzmann velocity distribution. For  finite 
$N$, relaxation to  equilibrium proceeds in two steps.  First, the system
relaxes to a quasi-stationary state (qSS), in which it stays for time $\tau_\times(N)$, 
after which it crosses over to the normal
thermodynamic equilibrium with the Maxwell-Boltzmann (MB) velocity distribution~\cite{Kav2007}. 
In the limit $N \rightarrow \infty$,
the life time of qSS diverges, $\tau_\times \rightarrow \infty$, and the thermodynamic
equilibrium is never reached.

Unlike the equilibrium state, which only depends on the global invariants such as the total 
energy and momentum
and is independent of the specifics of the initial particle  distribution,
the SS explicitly depends  on the initial condition.  This is the case for self-gravitating 
systems~\cite{Pa90}, 
confined one component plasmas~\cite{Levinprl,Rizz09},  geophysical systems~\cite{Cha05}, 
vortex dynamics~\cite{Miller90,Cha96,Venaille09}, etc~\cite{Campa09}, for which 
the SS state often has a  peculiar core-halo structure~\cite{Levinprl}.
In the thermodynamic limit, none of these systems 
can be described by the usual equilibrium statistical mechanics, and new methods must be developed.

In this paper we will restrict our attention to self-gravitating systems.  
Unfortunately, it is very hard to study  these systems in 3d~\cite{Levingrav,Joyce09}.  
The reason for this is that the 3d Newton potential is not
confining.  Some particles can gain enough energy to completely escape from 
the gravitational cluster, going all the way
to infinity.
In the thermodynamic limit, one must then consider three distinct populations: 
particles which will relax to form the central core,
particles which will form the halo, and  particles which will completely evaporate.  Existence
of three distinct classes of particles, makes the study of 3d systems particularly difficult.  
On the other hand, the interaction potential in 2d is logarithmic, 
so that all the particles remain gravitationally bound.  
Similar to  magnetically confined plasmas
the stationary state of a 2d gravitational system should, therefore, have a core-halo structure~\cite{Levinprl}. 
We thus expect
that the  insights gained from the study of confined plasmas might  prove to be useful to understand the 2d 
gravitational systems.  

\section{The Model}

Our system  consists of $N$ particles with the total mass $M$ in a two dimensional space.  At $t=0$ 
the particles are distributed
over the phase space with the initial distribution $f_0({\bf r},{\bf v})$, and then allowed to relax. 
Our goal is to calculate the one particle distribution function $f({\bf r},{\bf v})$, 
once the relaxation process has been completed and the stationary state has been established.
For now,
we will restrict our attention to  the azimuthally symmetric systems. 

The mean gravitational potential at $r$  at time $t$ satisfies the Poisson
equation 
\begin{equation}
\nabla^2\psi=4\pi Gm n({\bf r};t) \label{poisson0}
\end{equation}
where $m=M/N$, $n({\bf r};t)=N\int{f({\bf r},{\bf v};t)}d^2{\bf v}$ is the particle number density,
and $G$ is the 
gravitational constant. It is convenient to define dimensionless variables by scaling 
lengths, velocities, potential, and energy to $L_0$ (arbitrary length scale), $V_0=\sqrt{2GM}$, $\psi_0=2G M$ 
and $E_0=M V_0^2=2 G M^2$, respectively.  In 3d space, our system corresponds to 
infinitely long parallel rods of line density $m$ interacting
through a pair  potential $\phi(r)=2Gm^2\ln(r)$.

\section{The Thermodynamic Equilibrium: Finite $N$}

If the system has a finite number of particles, after sufficiently large time $\tau_\times(N)$, it will relax
to thermodynamics equilibrium with the
MB  distribution function, given {\it exactly}  by
\begin{equation}
\label{eqMB}
f_{MB}=Ce^{-\beta({\bf v}^{2}/2+\omega({\bf r}))}
\end{equation}
where $C$ is a normalization constant, $\beta=1/T$ is the Lagrange multiplier
used to conserve the total energy, and  $\omega({\bf r})$ 
is the potential of mean force~\cite{Levinrep}.  For  a gravitational system of mass $M$, the correlations
between the particles vanish as $N$ becomes large, so that  $\omega({\bf r}) \approx \psi({\bf r})$.  
Substituting Eq.~(\ref{eqMB}) in 
Eq.~(\ref{poisson0}), we obtain the classical Poisson-Boltzmann equation
in its adimensional form
\begin{equation}
\label{eqPB}
\nabla^2\psi=\frac{4\pi^2C}{\beta}e^{-\beta \psi}.
\end{equation}
The solution of this equation is~\cite{Ostriker64}
\begin{equation}
\label{pot PB}
\psi(r)=\frac{2}{\beta}\ln\left(\lambda^2+\frac{\pi^2C}{2\lambda^2}r^2\right).
\end{equation}
For large $r$ this potential must grow as
\begin{equation}
\label{gausscond}
\lim_{r \rightarrow \infty}[{\psi(r)}-\ln(r)]=0 \,,
\end{equation}
which requires that $\beta=4$ and  $\lambda^2=\pi^2C/2$. With these values, the distribution 
function Eq. (\ref{eqMB}) automatically satisfies the constraint
\begin{equation}
\label{eq conserv1}
\int{\rm d^2{\bf r}}\,{\rm d^2{\bf v}} f({\bf r},{\bf v})=1 \,,
\end{equation}
while the value of $\lambda$ is obtained from the conservation of energy
\begin{eqnarray}
\label{conserv2}
\int{\rm d^2{\bf r}}\,{\rm d^2{\bf v}}\;\left[\frac{{\bf v}^2}{2}+
\frac{\psi(r)}{2}\right] f({\bf r},{\bf v})={\cal E}_0  \,,
\end{eqnarray}
where ${\cal E}_0$ is the renormalized initial energy, see Appendix~\ref{energy}. 
In this paper we will restrict our attention to initial distributions of the water-bag form,
\begin{eqnarray}
\label{eq waterbag}
f_0({\bf r},{\bf v})=\eta \Theta(r_m-r)\Theta(v_m-v)\;,
\end{eqnarray}
where $\Theta(x)$ is the Heaviside step function and $\eta=1/\pi^2 r_m^2 v_m^2$.  
For simplicity, from now on we will measure all lengths in units of $r_m$, so that $r_m=1$.  
The renormalized energy, Appendix~\ref{energy}, then reduces to
\begin{equation}
{\cal E}_0=\frac{v_m^2}{4}-\frac{1}{8}\,.
\end{equation}
Performing the integral in Eq.~(\ref{conserv2}) with $\psi(r)$ given by Eq.~(\ref{pot PB}), we obtain
\begin{equation}
\lambda^2=e^{2(2{\cal E}_0-1)}\,.
\end{equation}
This provides a complete solution to the equilibrium 
thermodynamics of 2d self-gravitating system in a large (but finite) $N$ limit. 
We next compare  the analytical solution presented above with the full $N$-body molecular dynamics 
simulation~\cite{Levingrav}.  
To do this we calculate
the number density of particles  between $[r,r+dr]$
\begin{equation}
\label{eqnr}
N(r)=2 \pi N r \int{\rm d^2{\bf v}} f_{MB}({\bf r},{\bf v})=\frac{2N\lambda^2r}{(\lambda^2+r^2)^2}
\end{equation}
and the number density of particles with velocity between $[v, v+dv]$, 
\begin{equation}
\label{eqnv}
N(v)=2 \pi N v \int{\rm d^2{\bf r}} f_{MB}({\bf r},{\bf v})=4Nve^{-2v^2}\,.
\end{equation}

\begin{figure}[!ht]
\begin{center}
\includegraphics[width=9cm]{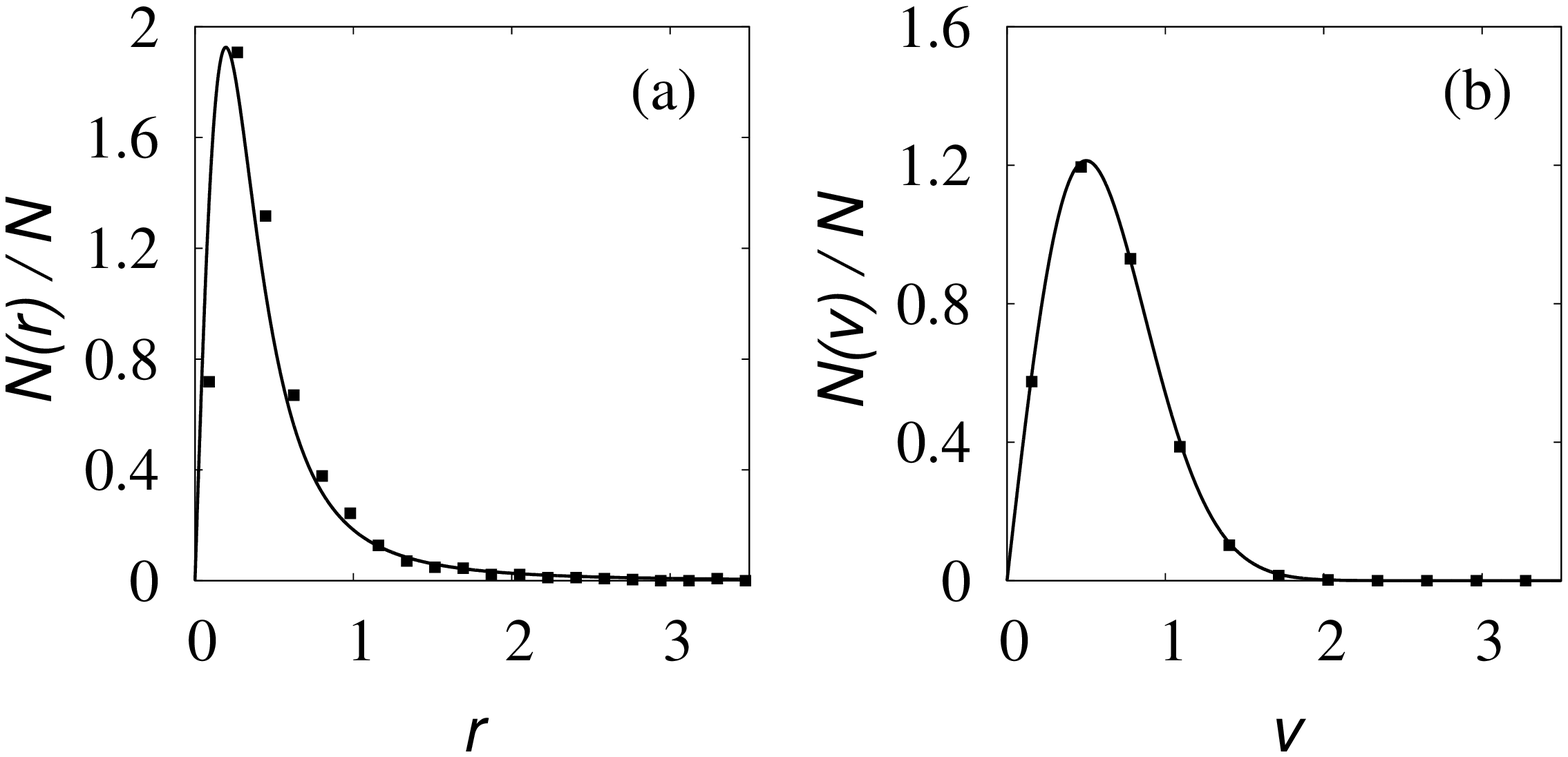}
\end{center}
\psfrag{r}{$r/\sqrt{K/\kappa_z}$}
\caption{(a) Position  and (b) velocity  distributions for a  system with 
${\cal E}_0=-0.0433673$.
Solid line is the theoretical prediction obtained using the MB distribution 
function, Eq.~(\ref{eqMB}), and the points are the result of molecular 
dynamics simulation with $N=10000$ particles.}
\label{fig-1}
\end{figure}
Fig. \ref{fig-1} shows an excellent agreement between the theory and the simulations.
It is important to stress, however,
that to reach the MB equilibrium distribution,  
has required a week of  CPU time, (a million dynamical times for $N=10000$ particles, see Section \ref{core-halo}).
Up to the crossover time $\tau_\times(N)$, 
the system remained trapped in a quasi-stationary state, with the one particle distribution
very different from the equilibrium one.  We now turn to the discussion of this non-equilibrium quasi-stationary state.

\section{The Thermodynamic Limit} 

For systems interacting through short-range potentials, the final stationary state corresponds to the
thermodynamic equilibrium and is exactly described by the MB distribution.  
In spite  of a popular believe that in the thermodynamic limit for systems with long-range unscreened
interactions the mean-field becomes exact, this is not quite true.  Or rather, this is
true mathematically, but is irrelevant for real physical systems, since when $N \rightarrow \infty$ 
it takes an infinite time for such a system to relax to the thermodynamic equilibrium. 
What is correct, is that in the thermodynamic limit the dynamical evolution of a system with 
long-range interactions 
is governed {\it exactly} by the
collisionless Boltzmann (Vlasov) equation~\cite{Br77}
\begin{equation}
\label{eq vlasov}
\frac{{\rm D}f}{{\rm D}t} \equiv {\partial f \over \partial t} + {\bf v} \cdot \nabla f 
+ \frac{{\bf F}}{m} \cdot \nabla_{\bf v} f =0, \\
\end{equation}
where $f$ is the one particle distribution function and ${\bf F}$ is the mean force felt by a particle at 
position $r$.  The MB distribution, together with the Poisson equation for the mean-field 
potential, is a stationary solution of the Vlasov equation.  Thus, if we start with this distribution it is 
guaranteed to
be preserved by the Vlasov dynamics.  However,  unlike for the collisional Boltzmann equation,  MB distribution 
is not a 
global attractor of the Vlasov dynamics --  an arbitrary (non-stationary) initial distribution will not evolve 
to the MB distribution.  Thus, the collisionless relaxation described by the Vlasov equation is much more complex than
the collisional relaxation governed by the  Boltzmann equation  for systems with short-range interactions.
The final stationary state of Vlasov dynamics will depends explicitly on the initial particle distribution.

Vlasov equation  has an infinite number of conserved quantities, called Casimir 
invariants.  Any local functional of the distribution function is a Casimir invariant 
of the Vlasov dynamics.
In particular, if we discretize the initial distribution function into surface levels with values $\{\eta_j\}$, 
the hypervolume
$\xi(\eta_j)=\int \delta(f({\bf r},{\bf v}; t)-\eta_j) {\rm d^d{\bf r}}{\rm d^d{\bf v}}$ of each level will be 
preserved by the Vlasov flow.  
The evolution of the distribution function corresponds to the process of
filamentation and proceeds {\it ad infinitum} from large to small length scale.  Thus on a fine-grain scale,
the evolution never stops and the stationary state is never reached.  However, in practice, there is always
a limit to the maximum resolution, and only a coarse-grained  distribution function is available in simulations or
in experiments.  It is this coarse-grained distribution which appears in practice as the stationary state of a 
collisionless relaxation dynamics. 

Numerical solution of the Vlasov equation is a very difficult task. Since 1960's there has been a
tremendous effort to find an alternative way to predict the final 
stationary state without having to explicitly solve the Vlasov 
equation~\cite{Ly67,Pa90,Ruffo95,Marciano09,Levinprl,Levingrav}.  One of the first statistical
approaches was proposed by Lynden-Bell
and has become know as the {\it violent relaxation theory} .
This theory is based on the assumption that there exists an efficient phase space mixing 
during the dynamical evolution. This assumption is similar to the ergodicity of the Boltzmann-Gibbs statistical
mechanics.   For systems with short-range interactions, 
ergodicity --- although very difficult to prove explicitly --- is almost always found to be satisfied in practice. This, however,
is not the case for the efficient mixing hypothesis for systems with long-range interactions.  
In fact it was found that for most initial conditions, the phase space mixing is very
poor. For magnetically confined plasmas, efficient mixing was found to exist only
for very special initial conditions, and in general these systems relax to a stationary state very
different from the one predicted by the Lynden-Bell theory.    
Similarly, for 3d gravitational systems, violent
relaxation  theory was found to work only if the 
initial distribution satisfied the, so called, virial condition~\cite{Levingrav,Joyce09}. 
Otherwise strong {\it particle-density wave} interactions broke the ergodicity and resulted in a core-halo phase
separation.  

\subsection{Violent relaxation}

We first briefly review the violent relaxation theory. The basic assumption of this
theory is that during the temporal evolution, the system is able to
efficiently explore the whole of phase space.
To obtain the stationary (coarse-grained) distribution $\bar f({\bf r},{\bf v})$, the
initial distribution $ f_0({\bf r},{\bf v})$ is discretized into the $p$ levels, and the 
phase space is divided into macrocells of volume ${\rm d^d{\bf r}}\,{\rm d^d{\bf v}}$, which
are in turn subdivided into $\nu$ microcells, each of volume $h^d$, for a d-dimensional system. 
Since Vlasov dynamics is incompressible, $\rm D f/\rm D t=0$, each microcell can contain
at most one discretized level $\eta_j$. The number density of the level $j$ 
inside a {\it macrocell} at $({\bf r},{\bf v})$ ---
number of microcells occupied by the level $j$ divided by $\nu$ --- 
will be denoted by $\rho_j({\bf r},{\bf v})$.
%
%
\begin{figure}[!h]
\vspace{1.cm}
\begin{minipage}{0.4\textwidth}
\begin{center}
\begin{pspicture}(2.5,2.5)
\psgrid[gridlabels=0pt,subgriddiv=0,gridwidth=1.25pt,subgridwidth=1.pt,subgriddots=4,subgridcolor=black]
(2.,2.)(0,0)(4,4)
\rput[b]{0}(4.,4.1){(a)}
\psframe*[linecolor=black](1.,1.)(3.,3.)
\psline{<->}(-0.1,0)(-0.1,1)
\rput[b]{0}(-0.43,0.43){d{\bf r}}
\psline{<->}(0,-0.1)(1,-0.1)
\rput[b]{0}(0.43,-0.43){d{\bf v}}
%
%
%
%

\psbezier[linewidth=1pt,showpoints=false]{->}(1.15,2.15)(0.48,2.48)(0.15,1.48)(-1,2.5)

\rput[b]{0}(-1.1,2.6){$\eta$}
\end{pspicture}
\end{center}
\end{minipage}
\begin{minipage}{0.4\textwidth}
\vspace{.5cm}
\hspace{-0.5cm}
\begin{center}
\begin{pspicture}(2.5,2.5)
\psgrid[gridlabels=0pt,subgriddiv=3,gridwidth=1.25pt,subgridwidth=1.pt,subgriddots=4,subgridcolor=black]
(2.,2.)(0,0)(4,4)
\rput[b]{0}(4.,4.1){(b)}
\psframe*(0,0)(0.34,0.34)
\psframe*(0.34,0.68)(0.68,1.)
\psframe*(0.68,0.34)(1.,0.68)
\psframe*(1,0)(1.34,0.34)
\psframe*(1.34,3.68)(1.68,4.)
\psframe*(1.0,3.)(1.34,3.34)
\psframe*(1.68,3.)(2.,3.34)
\psframe*(1.34,2.68)(1.68,3.)
\psframe*(3,3)(3.34,3.34)
\psframe*(2.34,2.34)(2.68,2.68)
\psframe*(1.68,1.68)(2.,2.)
\psframe*(1.34,1.34)(1.68,1.68)
\psframe*(1.68,1.)(2.,1.34)
\psframe*(2.34,1.68)(2.68,2.)
\psframe*(3.,1.68)(3.34,2.)
\psframe*(3.34,1.34)(3.68,1.68)
\psframe*(1.34,.68)(1.68,1.)
\psframe*(2.68,2)(3.,2.34)
\psframe*(2,1.34)(2.34,1.68)
\psframe*(1,2)(1.34,2.34)
\psframe*(0.68,1.68)(1.,2.)
\psframe*(0.34,2.68)(0.68,3.)
\psframe*(0.34,3.34)(0.68,3.68)
\psframe*(0,3.68)(0.34,4.)
\psframe*(2,3.34)(2.34,3.68)
\psframe*(2.68,3.68)(3.,4.)
\psframe*(2.34,3)(2.68,3.34)
\psframe*(3.34,3.68)(3.68,4.)
\psframe*(3.68,2)(4.,2.34)
\psframe*(3.34,2.34)(3.68,2.68)
\psframe*(3.68,2.68)(4.,3.)
\psframe*(2.34,0.34)(2.68,0.68)
\psframe*(2,0)(2.34,0.34)
\psframe*(3.,0.68)(3.34,1.)
\psframe*(3.34,0.34)(3.68,0.68)
\psframe*(3.68,0)(4.,0.34)

\psline{<->}(-0.1,0)(-0.1,1)
\rput[b]{0}(-0.43,0.43){d{\bf r}}
\psline{<->}(0,-0.1)(1,-0.1)
\rput[b]{0}(0.43,-0.43){d{\bf v}}

\psline{<->}(-0.1,3.68)(-0.1,4)
\rput[b]{0}(-0.4,3.75){\it h}

\pnode(4,3){A}
\pnode(4,4){B}
\ncarc[arcangle=-90]{<->}{A}{B}
\rput[b]{0}(5.2,3.4){Macro-cell}
\rput[b]{0}(5.2,2.9){$\bar f$({\bf r},{\bf v})}

\pnode(4,1.68){C}
\pnode(4,2.){D}
\ncarc[arcangle=-90]{<->}{C}{D}
\rput[b]{0}(5.2,1.75){Micro-cell}
\rput[b]{0}(5.2,1.25){$f$({\bf r},{\bf v})}

\psbezier[linewidth=1pt,showpoints=false]{->}(1.15,2.15)(0.48,2.48)(0.15,1.48)(-1,2.5)
\psbezier[linewidth=1pt,showpoints=false]{->}(2.45,3.15)(1.48,3.48)(1.15,2.48)(-1,2.6)
\rput[b]{0}(-1.1,2.45){$\eta$}
\end{pspicture}
\vspace{0.75cm}
\end{center}
\end{minipage}
\caption{Coarsening of phase-space described by the Vlasov dynamics: (a) initial and (b) final 
stationary state for a
distribution with initial phase-space density $\eta$. In this example,  
$d=1$, $p=1$ and $\nu=9$.}
\label{fig0}
\end{figure}
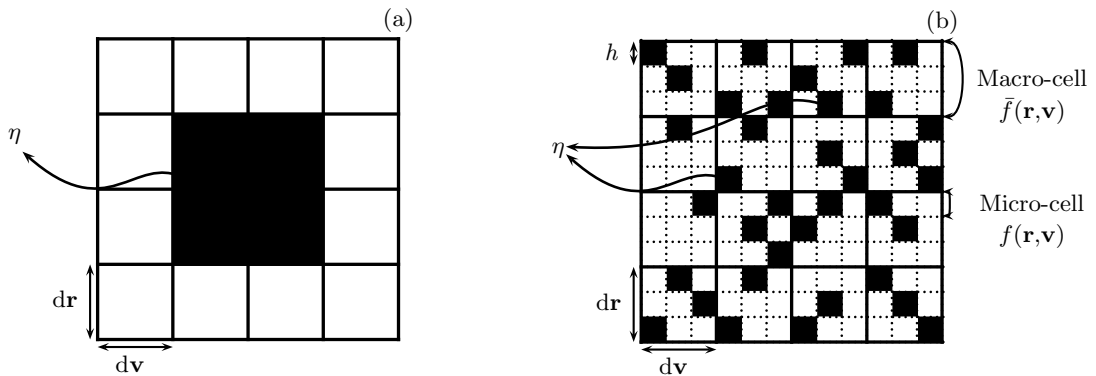
%
%
Note that by construction,
the total number density of {\it all} levels in a macrocell 
is restricted to be 
\begin{equation}
 \label{eq vol}
\sum_j \rho_j({\bf r}, {\bf v}) \le 1 \;,
\end{equation}
see Fig.~\ref{fig0}.
Using a standard combinatorial procedure~\cite{Ly67,Levinprl} it is then possible to associate
a coarse-grained entropy with the distribution of $\{\rho_j\}$. 
The entropy is found to be that of a $p$-species lattice gas,
\begin{eqnarray}
\label{eq entropy}
S&=&-\int \frac{{\rm d^d{\bf r}}\,{\rm d^d{\bf v}}}{h^d}  \left\{ \sum_{j=1}^p \rho_j ({\bf r},{\bf v}) 
\ln[\rho_j ({\bf r},{\bf v})]
 + [1-\sum_{j=1}^p\rho_j  ({\bf r},{\bf v})] \ln[1-\sum_{j=1}^p\rho_j ({\bf r},{\bf v})] \right\} \;,
\end{eqnarray}
with  the Boltzmann constant  set to one.
If the initial condition is of the water-bag form, Eq.~(\ref{eq waterbag}) ($p=1$),
the maximization procedure
is particularly simple, yielding a Fermi-Dirac distribution,
\begin{eqnarray}
\label{eq lynbell}
\bar f({\bf r},{\bf v})=\eta \rho({\bf r},{\bf v})=
\frac{\eta}{e^{\beta [\epsilon({\bf r},{\bf v})-\mu]}+1}\;,
\end{eqnarray}
where $\epsilon({\bf r}, {\bf v})=\frac{1}{2} {\bf v}^2 + \psi(r)$ is the mean particle energy,  and 
$\beta$ and $\mu$ are the two Lagrange multipliers required by the conservation of  the total number of particles and 
the total energy Eqs.~(\ref{eq conserv1}) and (\ref{conserv2}).

\section{Virial Cases}
\label{virial}
For a 2d self-gravitating system the virial 
theorem requires that $\langle v^2 \rangle =1/2$, in a stationary state Appendix~\ref{Virial}.
If the initial distribution does not satisfy this condition, the system will undergo strong oscillations 
before relaxing into the final stationary state in which the virial theorem is satisfied.  
For a water-bag initial distribution the virial condition reduces  to the requirement that  $v_m=1$.
For future convenience, we will define the virial number for water-bag distributions to be $\mu\equiv 1/v_m$,
so that $\mu=1$, when the initial distribution satisfies the virial condition.  If $\mu \ne 1$, 
the envelope radius, defined as
$r_e(t)=\sqrt{2 \langle r^2 \rangle}$, will vary with time  until a stationary state is achieved. 
Note that with the above definition, $r_e(0)=1$, as it should be. 
It is possible to show that the temporal
evolution of  $r_e(t)$ satisfies
\begin{equation}
\ddot{r}_e(t)+\frac{1}{r_e(t)}=\frac{\varepsilon^2(t)}{r_e^3(t)}\;, \label{envelope0}
\end{equation}
where  $\varepsilon^2=4\left
[\langle {\bf r}^2 \rangle\langle {\dot{\bf r}}^2 \rangle - 
\langle {\bf r}\cdot\dot{{\bf r}}\rangle^2\right]$. The derivation of this equation is given in
Appendix~\ref{Envelope}.  For an initial water-bag distribution, 
$\langle {\bf r}(0)\cdot\dot{{\bf r}}(0)\rangle=0$ and $\langle {\bf r}^2(0) \rangle\langle 
{\dot{\bf r}}^2(0) \rangle=v_m^2/4$
so that, if the initial distribution satisfies the virial condition, $v_m=1 \Rightarrow \ddot{r}_e=0$, and the large
envelope oscillations are suppressed.  
As was already noted for magnetically confined plasmas and 3d self-gravitating
systems~\cite{Levinprl,Levingrav}, we expect the violent relaxation theory to work well
when the initial distribution satisfies the virial condition and there are no macroscopic 
envelope oscillations.  To check this, we compare the predictions of the theory with the full
$N$-particle molecular dynamics simulations. 
At $t=0$,  particles are distributed over the phase space in 
accordance with 
the water-bag distribution~(\ref{eq waterbag}) which satisfies the 
virial condition, $\mu=1$.  
\begin{figure}[!htb]
\begin{center}
\includegraphics[width=9cm]{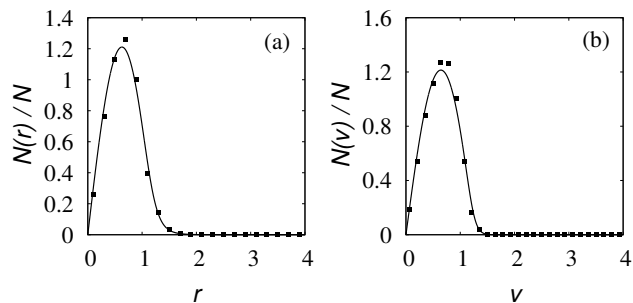}
\end{center}
\caption{Position (a) and velocity (b) distributions for a system satisfy the  virial condition. 
The solid line is the theoretical prediction obtained using the distribution function of Eq.~(\ref{eq lynbell}), 
while the points are the result of molecular dynamics simulation with $N=10000$ particles.}
\label{fig1}
\end{figure}
We then numerically solve the Poisson equation (\ref{poisson0}), with the distribution function given
by equation (\ref{eq lynbell}) and compare the results with the molecular dynamics simulations.
As can be seen from Fig.~\ref{fig1} there is a reasonably good agreement between the theory
and the simulations. 
\begin{figure}[!ht]
\begin{center}
\includegraphics[width=6cm]{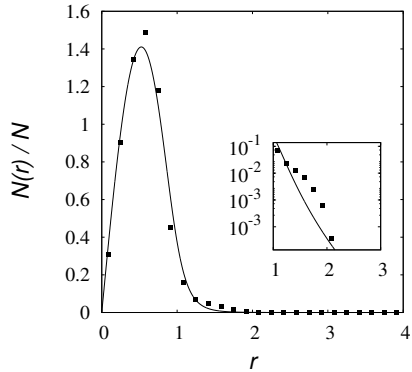}
\end{center}
\caption{(a) Position distribution for a nearly virial self-gravitating system with initial $\mu=1.2$. 
Solid line is the theoretical prediction obtained using the distribution function of Eq.~(\ref{eq lynbell}), 
while the points are the result of molecular dynamics simulation with $N=10000$ particles.}
\label{fig2}
\end{figure}
However, if the virial condition is not met exactly, one  notices a deviation in the tail region of the
particle distribution, see inset of  Fig.~\ref{fig2}. 
For initial distributions with $\mu$ significantly different from $1$, there is a clear qualitative change
in the SS distribution function.  In this case 
the original homogeneous cluster, separates into a high density core region
surrounded by a diffuse halo, Fig.~\ref{fig3} ---  the 
violent relaxation theory fails completely and a new approach must be
developed~\cite{Levinprl,Levingrav,Teles09}.
\begin{figure}[!ht]
\begin{center}
\includegraphics[width=9cm]{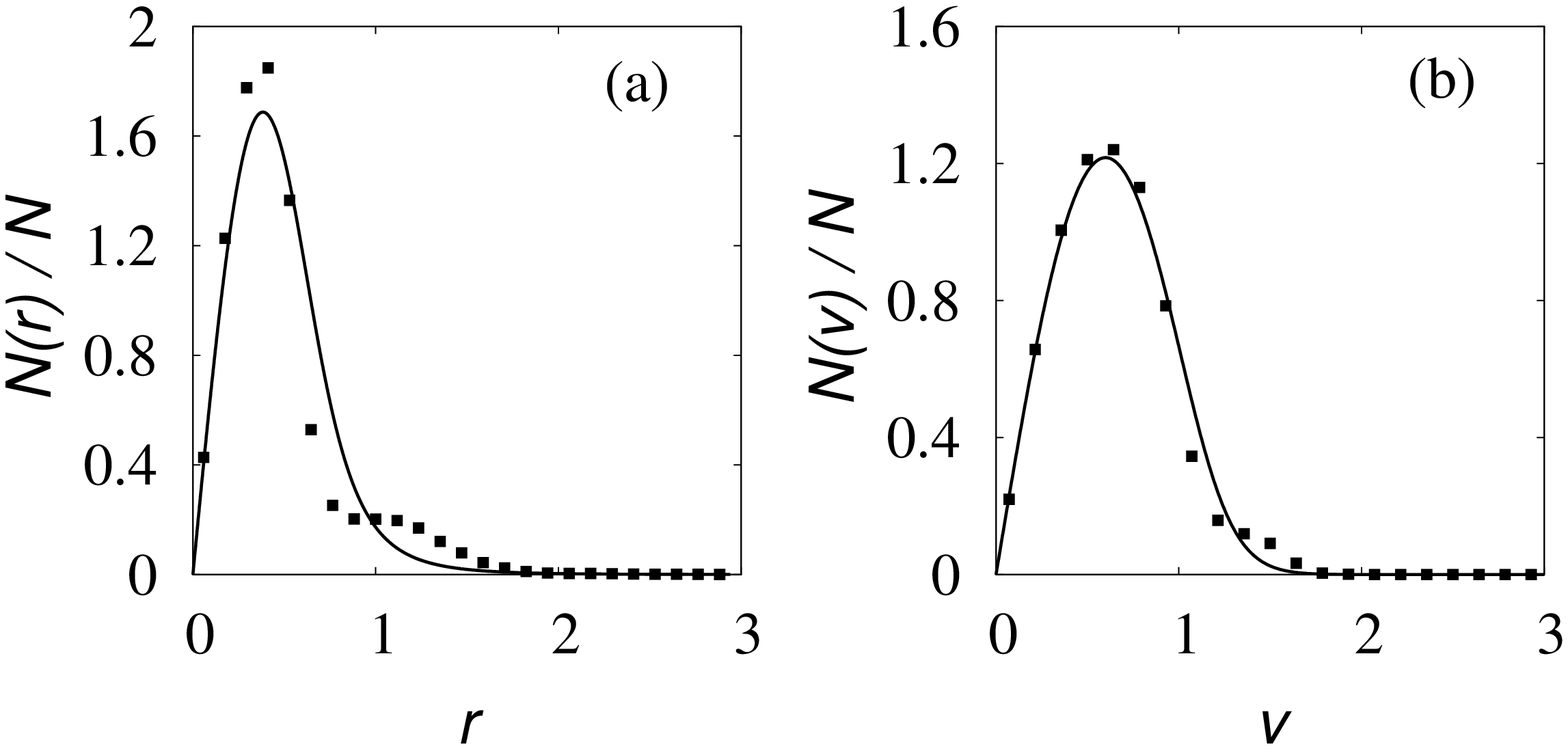}
\end{center}
\vspace{-0.5cm}
\caption{Position (a) and velocity (b) distributions for a  system with $\mu=1.7$. 
Solid line is the prediction of the violent relaxation theory 
Eq.~(\ref{eq lynbell}), while points are the result of molecular dynamics simulation with $N=10000$ particles.}
\label{fig3}
\end{figure}

\section{Core-halo Distribution}

\begin{figure}[!htb]
\begin{center}
\includegraphics[width=9cm]{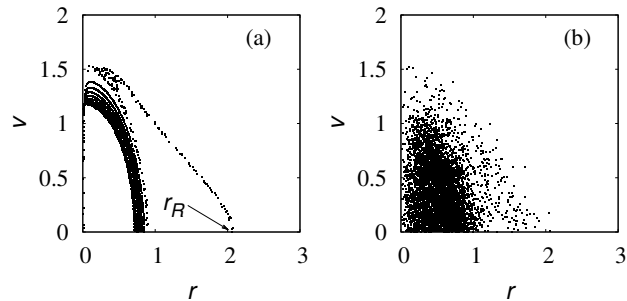}
\end{center}
\vspace{-0.5cm}
\caption{(a) Poincar{\'e} plot of a test particle in a oscillating 
potential, Eq.~(\ref{eqpoinc}) with 10 different initial conditions, 
plotted when the envelope is at its minimum. 
(b) N-particle simulation for a non-virial system with $\mu=1.2$. 
An excellent agreement is found between the extent of the halo in N-particle simulation and the 
one particle resonant orbit shown in the Poincar{\'e}
plot. }
\label{fig4}
\end{figure}

The failure of the violent relaxation theory is a consequence of the inapplicability of the efficient
mixing hypothesis to strongly oscillating  gravitational systems, Figs.~\ref{fig2} and~\ref{fig3}. 
Density oscillations
excite parametric resonances which favor some  particles to gain a lot of energy at the expense 
of the rest.  The resulting particle-wave interactions are a form of  non-linear Landau damping  
which allows some particles to escape from the main cluster to form a diffuse halo.  
The process of evaporation will continue as long as the oscillations of the core persist. Oscillations will 
only stop
when the core  exhaust all of its free energy, and its effective temperature drops to  
$T \approx 0$, $\beta \rightarrow \infty$ in Eq. (\ref{eq lynbell}). Note that 
because of the incompressibility restriction 
imposed by the Vlasov dynamics Eq. (\ref{eq vol}), the
core can not freeze -- collapse to the minimum
of the potential energy. Instead, the distribution function of
the core particles progressively approaches that of a fully
degenerate Fermi gas~\cite{Levingrav},
\begin{eqnarray}
\label{eq core}
\bar f_{core}\left({\bf r},{\bf v}\right)&=&\eta\Theta\left(\epsilon_F-\epsilon({\bf r},{\bf v})\right)
\end{eqnarray}
where $\epsilon_F$ is the effective Fermi energy.
The final stationary state of the cluster will then correspond to 
a cold core surrounded by a high energy diffuse halo, 
\begin{eqnarray}
\label{eq corehalo}
\bar f\left({\bf r},{\bf v}\right)&=&\eta\Theta\left(\epsilon_F-\epsilon({\bf r},{\bf v})\right)
+\chi\Theta(\epsilon({\bf r},{\bf v})-\epsilon_F)\Theta(\epsilon_R-\epsilon({\bf r},{\bf v}))\,,
\end{eqnarray}
where $\epsilon_R$ is the energy of the one particle resonance.
The parameter $\chi$ and the effective Fermi
energy $\epsilon_F$ are determined using the conservation of particle number and  energy. 
The extent and the location of
the parametric resonance can be calculated
using the canonical perturbation theory~\cite{Gluck94}.
In Fig. \ref{fig4} we show the Poincar\'{e} section of a test particle $i$ moving
under the action of an oscillating potential calculated
using the envelope equation~(\ref{envelope0}),
\begin{eqnarray}
\label{eqpoinc}
\ddot{r}_i(t)-\frac{L_i^2}{r_i^3(t)}&=&
\left\{
\begin{array}{l}
-\frac{r_i(t)}{r_e^2(t)}\>\, \text{ for }\>\,r_i(t)\le r_e(t)
\\
\\
-\frac{1}{r_i(t)}\>\, \text{ for }\>\,r_i(t)\ge r_e(t)
\end{array}
\right.
\end{eqnarray}
with,
\begin{equation}
\ddot{r}_e(t)+\frac{1}{r_e(t)}=\frac{\varepsilon^2_0}{r_e^3(t)}\;, \label{eqenvelop}
\end{equation}
where $L_i=|{\bf r}_i\times{\bf v}_i|$ is the modulus of the  test particle angular momentum, conserved 
by the dynamics, and $\varepsilon(t)$ is fixed at its initial value $\varepsilon(t)=\varepsilon_0=v_m=1/\mu$.
The resonant orbit is
the outermost curve of the Poincar\'e plot Fig.~\ref{fig4}. The first resonant
particles move in an almost  simple harmonic motion with  energy
$\epsilon_R=\ln(r_R)$, where $r_R$ is the intersection of the 
resonant trajectory with the $v=0$ axis. 

Empirically we find that the  location 
of the one particle resonance $r_R$ for values of $|\mu-1|>0.1$
is very well approximated by a simple  expression~\cite{Wang98} 
\begin{equation}
r_R=2\left(1+|\ln \mu \right|)/\mu\;.
\end{equation}
As the relaxation proceeds, the oscillating core
becomes progressively colder, while a halo of highly energetic 
particles is formed. 
As more and more particles are ejected from the  core,
their motion becomes chaotic, and a halo distribution becomes smeared out. 
Similar to what happens for magnetically confined plasmas,  
we find that the distribution function of a completely relaxed halo is 
very well approximated by the Heaviside step function $\Theta(\epsilon_R-\epsilon({\bf r},{\bf v}))$.
For notational simplicity, from now on, we will drop the over-bar on the distribution function
$f\left({\bf r},{\bf v}\right)$, but it should always be kept in mind that $f$ is stationary only within
the coarse graining procedure described above.  

\section{Analytical solution to the core-halo problem}
\label{core-halo}
 
In order to obtain the density and the velocity distribution after the SS state is achieved,
we solve the Poisson equation~(\ref{poisson0})
\begin{equation}
\nabla^2 \psi({\bf r}) = 2\pi \int{f\left({\bf r},{\bf v}\right) d^2{\bf v}}, \label{eq poisson}
\end{equation}
with the constraints~(\ref{eq conserv1}) and (\ref{conserv2}). 
Since the initial mass distribution has the azimuthal symmetry,
the potential must have the form
\begin{eqnarray} 
\psi(r)&=&\psi_{core}(r)\Theta\left(r_c-r\right)+\psi_{halo}(r)\Theta(r-r_c)\Theta(r_R-r)+\psi_{out}(r)\Theta(r-r_R).
\end{eqnarray}
Substituting this into Poisson equation and noting that $\epsilon_F=\psi(r_c)$ we obtain
\begin{eqnarray}
\psi_{core}(r)&=&\epsilon_R+C_1[(\eta/\chi-1)J_0(r_c^*)+J_0(r^*)]\\
\psi_{halo}(r)&=&\epsilon_R+C_2\>J_0(r^{**})+C_3\>Y_0(r^{**})\\
\psi_{out}(r)&=&\ln \left(r\right) \;,
\end{eqnarray}
where $J_0$ and $Y_0$ are the Bessel functions of the first type and of order $0$; $r_c$ is the core radius; 
$r^{*}=2\pi r \sqrt{\eta}$ and $r^{**}=2\pi r \sqrt{\chi}$. 
The integration constants $C_{1,2,3}$ and the value of $r_c$ can be determined using the
continuity of the potential and of the gravitational field.  The parameter $\chi$ can then be obtained
using the conservation of energy.   Once   $C_{1,2,3}$ are calculated, see  Appendix~\ref{Constants}, 
we are left with just two equations for $r_c$ and $\chi$,
\begin{eqnarray}
\left\{
\begin{array}{l}
{\cal E}(r_c,\chi)-{\cal E}_0 = 0 \\
\psi'_{core}(r_c)=\psi'_{halo}(r_c) \,,
\end{array}
\right.
\label{econ}
\end{eqnarray}
where prime denotes the derivative with respect to $r$ and
\begin{eqnarray}
{\cal E}(r_c,\chi)&=&\frac{\epsilon_R}{2}-\frac{\pi^4\chi(\eta -\chi )
J_2\left(r_c^{*}\right)r_c^2 \left[Y_0 \left(r_c^{**}\right)J_0\left(r_R^{**}\right)-
J_0\left(r_c^{**}\right)Y_0\left(r_R^{**}\right)\right]^2}{4\eta J_0\left(r_c^{*}\right)}\;.\\
\end{eqnarray}

\begin{figure}[!hb]
\begin{center}
\includegraphics[width=9cm]{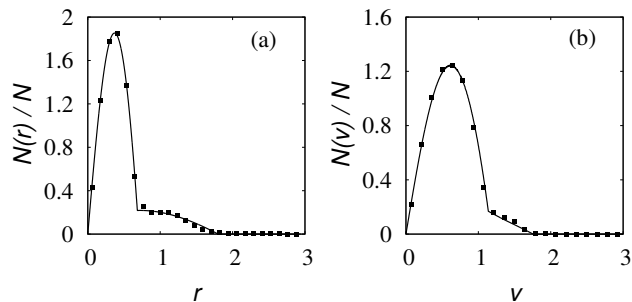}
\end{center}
\caption{Position (a) and velocity (b) distributions for a  system with $\mu=1.7$.
Solid line is the theoretical prediction obtained using the distribution 
function, Eq.~(\ref{eq corehalo}), and the points are the result of molecular dynamics simulation with $N=10000$ 
particles.}
\label{fig5}
\end{figure}

This completely determines the distribution function of the final stationary state achieved
by a self-gravitating system when its initial distribution deviates from the virial condition.
In Fig. \ref{fig5}, we compare the predictions of the theory with the molecular dynamics
simulations.  An excellent agreement is found without any adjustable parameters.

\begin{figure}[!ht]
\begin{center}
\begin{minipage}{0.45\textwidth}
\hspace{-7.cm}
\includegraphics[width=5cm]{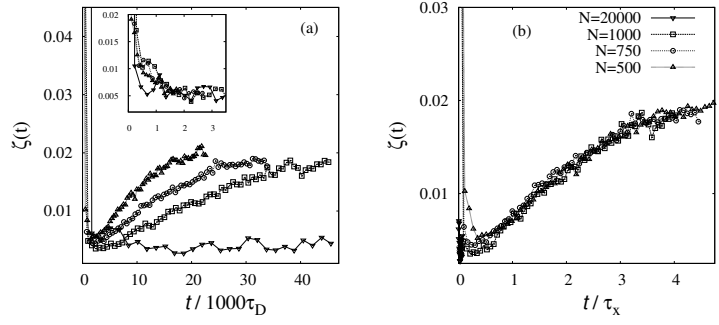}
\end{minipage}
\end{center}
\caption{(a) $\zeta(t)$ for different number of particles in the system.  After relaxing into the qSS,
the system  crosses over to MB distribution after time  $\tau_\times(N)$.
Inset (a) shows that the relaxation to core-halo state takes approximately 
$t\approx 2000\tau_D$ and does not depend on the number of particles in the system.
When the time is scaled with $\tau_\times(N)$ all the data in (a) (for large times) collapse
onto one universal curve (b).}
\label{fig7}
\end{figure}

Finally, we explore the life time $\tau_\times(N)$ of a qSS of a self-gravitating system with a
finite number of particles. To do this we define the crossover parameter
\begin{equation}
\zeta(t)=\frac{1}{N^2}\int_0^{\infty}{[N(r,t)-N_{ch}(r)]^2{\rm d}r}
\end{equation}
where $N(r,t)$ is the number density of particles inside shells located between $r$ and $r+{\rm d}r$ at
each time of simulation $t$ and $N_{ch}(r)=2\pi Nr\int{f_{ch}\left({\bf r},{\bf v}\right){\rm d}^2{\bf v}}$ 
where $f_{ch}\left({\bf r},{\bf v}\right)$ is the stationary distribution given by 
Eq.~(\ref{eq corehalo}). The dynamical time scale is defined as $\tau_D=r_m/\sqrt{2GM}$.  
In Fig. \ref{fig7}a we plot the value of $\zeta(t)$ for systems with different number of particles. Fig. \ref{fig7}b
shows that if we scale the time with  $\tau_\times=N^{\gamma}\tau_D$, where $\gamma=1.35$, all the curves
collapse onto one universal curve, showing the divergence of the crossover time in the thermodynamic limit. It is interesting
to note that for a Hamiltonian Mean-Field (HMF) model,  $\tau_\times(N)$ was found to diverge with the 
exponent $\gamma=1.7$~\cite{Yama2004}, while for a virial 3d self-gravitating system the exponent was found 
to be $\gamma \approx 1$.
Unfortunately, at the moment there is no theory which allows us to predict this exponents {\it a priory}.

\clearpage

\section{Conclusions}

We have studied the thermodynamics of 2d self-gravitating system in the microcanonical ensemble.
It was shown that the gravitational clusters containing  finite number of particles relax to  
the equilibrium state
characterized by the MB distribution.  Prior to achieving the thermodynamic equilibrium, however, these
systems become trapped in a quasi-stationary state,  where they stay for time $\tau_\times$,
which diverges as $N^{1.35}$ for large $N$.  Thus, in the limit $N \rightarrow \infty$ at fixed total mass $M$,
thermodynamic equilibrium can not be reached in a finite time.  A new
approach, based of the conservation properties of the Vlasov dynamics and on the theory of parametric resonances,  
is formulated and allows us to quantitatively predict the one particle distribution function in the non-equilibrium 
stationary state. 
Finally, it is curious to consider what will happen to a self-gravitating system in a contact with a thermal bath --- 
the canonical ensemble.
In Appendix \ref{Virial}, it is shown that for a 2d self-gravitating system a stationary state is possible,
if and only if, $\langle v^2 \rangle$=1/2, i.e. when the kinetic temperature is $T=1/4$.  If such system is put into contact
with a thermal bath which has $T>1/4$ there will be a constant heat flux from the reservoir into the system.  
This heat will be converted into    
the gravitational potential energy --- since the kinetic energy is fixed by the virial condition ---  making the cluster
expand without a limit.  
Conversely if the bath temperature is  $T<1/4$, the heat flux will be from the system into the bath.
Again, since the system can only exist in a stationary state if $T=1/4$, the energy for the heat flux can 
come only from the gravitational
potential.  In this case the gravitational cluster will contract without a limit, concentrating all of 
its mass at the origin.  Thus,  in the canonical ensemble no thermodynamic equilibrium is possible, unless the reservoir is at exactly $T=1/4$.  
We hope that the present work will also help shed new
light on the collisionless relaxation in 3d self-gravitating systems.  Unfortunately the 
3d problem  is significantly more difficult, since besides the core-halo formation, one
must also account for the particles evaporating to infinity.

This work was partially supported by the CNPq, INCT-FCx, and by the US-AFOSR under the grant FA9550-09-1-0283.

\clearpage
\appendix
\section{The Total Energy}
\label{energy}
The gravitational potential energy $U$ of a system is
\begin{equation}
U=-\frac{1}{4\pi}\int{\left(-\nabla \psi\right)^2{\rm d^2{\bf r}}}
\end{equation}
where the integration extends over all space. Unlike in 3d, the gravitational  potential of a 2d system
diverges at infinity. Therefore, some care must be taken with the limits. Performing the 
integration by parts we obtain
\begin{equation}
U=\frac{1}{2}\int{\left(\psi(r)-\lim_{r_0\rightarrow \infty}{\psi(r_0)}\right) 
f({\bf r},{\bf v};t)\rm d^2{\bf r}}\,{\rm d^2{\bf v}} \,,
\end{equation}
where $r_0$ is the radius of the  bounding sphere.  From Eq.~(\ref{gausscond}), $\psi(r_0)=\ln(r_0)$,
and the total energy is given by
\begin{eqnarray}
\label{eq0}
E=\int{\left(\frac{{\bf v}^2}{2}+\frac{\psi(r)}{2}\right)
f({\bf r},{\bf v};t){\rm d^2{\bf r}}\,{\rm d^2{\bf v}}}-\frac{1}{2}\lim_{r_0 \rightarrow \infty}\ln(r_0) \\ \nonumber 
\end{eqnarray}
The divergence in the last term is common to all energy calculations in 2d.   For example, if 
at $t=0$ the system is distributed with a water-bag distribution Eq. (\ref{eq waterbag}), its energy is 
\begin{equation}
\label{eq ener0}
E_0=\frac{v_m^2}{4}-\frac{1}{8}+\frac{1}{2}\ln\left(r_m\right)-\frac{1}{2}\lim_{r_0 \rightarrow \infty}\ln(r_0)\,.
\end{equation}
The gravitational self-energy in 2d is always divergent.  However, since this divergence is always the
same, it can be easily renormalized away.  We simply add an infinite constant, 
$\frac{1}{2}\lim_{r_0 \rightarrow \infty}\ln(r_0)$, to all 
gravitational self-energies.   The renormalized (finite) 
energy ${\cal E}$ of a self-gravitating system is then 
\begin{eqnarray}
\label{eq1}
{\cal E}=\int{\left(\frac{{\bf v}^2}{2}+
\frac{\psi(r)}{2}\right) f({\bf r},{\bf v})\rm d^2{\bf r}}\,{\rm d^2{\bf v}} \\ \nonumber .
\end{eqnarray}
which for a  water-bag distribution Eq.~(\ref{eq waterbag}) becomes
\begin{equation}
{\cal E}_0=\frac{v_m^2}{4}-\frac{1}{8}+\frac{1}{2}\ln(r_m).
\end{equation}
\section{The Virial Theorem}
\label{Virial}
The Hamiltonian of a general self-confined system is given by
\begin{eqnarray}
{\cal H}=\sum_i\frac{{\bf p}_i^2}{2m}+\frac{1}{2}\sum_{i j}V({\bf r}_i-{\bf r}_j) \nonumber
\end{eqnarray}
where ${\bf p}_i$ is the momentum of particle $i$, and $V({\bf r}_i-{\bf r}_j)$ is the interaction 
potential.  The virial function $I$ is defined as
\begin{eqnarray}
I=\langle \sum_i{\bf r}_i{\bf p}_i \rangle\,. \nonumber
\end{eqnarray}
Taking the time derivative, and using Hamilton's equations we obtain
\begin{eqnarray}
\label{eq A1}
\frac{\rm d}{\rm dt}I&=&\langle \sum_i\frac{{\bf p}_i^2}{m} \rangle-
\langle \sum_{i}{\bf r}_i\frac{\partial}{\partial{\bf r}_i}\tilde V \rangle \nonumber\\
&=&\sum_i\langle \frac{{\bf p}_i^2}{m} \rangle-\langle \sum_{i}{\bf r}_i
\frac{\partial}{\partial{\bf r}_i}\tilde V \rangle
\end{eqnarray}
where,
\begin{eqnarray}
\tilde V=\frac{1}{2}\sum_{i j}V({\bf r}_i-{\bf r}_j).\nonumber
\end{eqnarray}
If $\tilde V$ is an homogeneous function of order $p$
\begin{eqnarray}
\tilde V({\bf r})=\lambda^{-p}\tilde V(\lambda{\bf r})\;, \nonumber
\end{eqnarray}
Euler's theorem requires that
\begin{eqnarray}
p\tilde V=\sum_i{\bf r}_i\frac{\partial}{\partial{\bf r}_i}\tilde V \nonumber
\end{eqnarray}
For a stationary state ${\rm d}I/{\rm dt}=0$, and we obtain the usual result $2{\cal K}=p\langle \tilde V \rangle$, 
where ${\cal K}$ is the mean kinetic energy. 

In two dimensions, $\tilde V$ is a sum of logarithms  and  Euler equation 
does not apply directly. However, we can still derive
a 2d virial theorem by  writing the inter-particle interaction 
potential as  $V({\bf r}) =2 G m^2 \lim_{p \rightarrow 0}(|{\bf r}|^p)/p$, which
is a logarithm plus an infinite constant.  This is a homogeneous function of order p=0, so we can use the
Euler theorem to write
\begin{eqnarray}
 G m^2 N(N-1)=\sum_i{\bf r}_i\frac{\partial}{\partial{\bf r}_i}{\tilde V}
\label{vir}
\end{eqnarray}
Note that since the right hand side of this equation only contains derivative of the potential,
this expression is equally valid for $p=0$ and for the logarithmic potential. 
Substituting Eq.~(\ref{vir}) into Eq.~(\ref{eq A1}), we arrive at 2d virial theorem~\cite{chav2006}.
\begin{equation}
\langle v^2 \rangle = G M\frac{N-1}{N}
\end{equation}
In the thermodynamic limit, and after rescaling the velocity to put everything into adimensional form, we obtain the
result $\langle v^2 \rangle=1/2$, quoted in Section~\ref{virial}. 

\clearpage
\section{The Envelope Equation}
\label{Envelope}
We define the $rms$ radius of the mass distribution as  $R\equiv\sqrt{\langle  {\bf r}^2  \rangle}$.
Deriving twice with respect to time we obtain
\begin{equation}
\ddot{R}=\frac{\langle  {{\bf r}}^2  \rangle\langle  {\dot{\bf r}}^2  \rangle}{R^3} - \frac{\langle  {\bf r}\cdot\dot{{\bf r}} 
\rangle^2}{R^3}+\frac{\langle  {\bf r}\cdot\ddot{{\bf r}}  \rangle}{R} \,.
\end{equation}
This reduces to
\begin{equation}
\ddot{R}=\frac{\varepsilon^2}{4 R^3} +\frac{\langle  {\bf r}\cdot\ddot{{\bf r}}  \rangle}{R}\label{envelope1}.
\end{equation}
where $\varepsilon^2\equiv 4\left(\langle  {\bf r}^2  \rangle\langle  {\dot{\bf r}}^2  
\rangle - \langle  {\bf r}\cdot\dot{{\bf r}} 
\rangle^2\right)$ is the 
{\it emittance}
which commonly appears in plasma  physics~\cite{Dav01}. The last term can be simplified using the 
Poisson equation~(\ref{poisson0}) in its dimensionless form,
\begin{eqnarray}
\langle {\bf r}\cdot\ddot{{\bf r}} \rangle&=&\int{{\bf r}\cdot\ddot{{\bf r}}\ f({\bf r},{\bf v},t)
d^2{\bf r}d^2{\bf v}}\nonumber\\
&=&\frac{1}{2\pi}\int{{\bf r}\cdot\ddot{{\bf r}}\ \nabla^2\psi d^2{\bf r}}\nonumber\\
&=&-\int{r^2{\partial\psi\over \partial r}\nabla^2\psi dr} \nonumber\\
&=&-\int{r{\partial\psi\over \partial r}{\partial\over\partial r}\left(r{\partial\psi\over \partial r}
\right)dr}\nonumber\\
&=&-\frac{1}{2}\int_0^{\infty}{dr {\partial\over\partial r}\left[\left(r\frac{\partial\psi}{\partial r}
\right)^2\right]}\nonumber\\
&=&\lim_{r_0 \rightarrow \infty}{-\frac{1}{2}\left(r{\partial\psi\over\partial r}\right)^2\bigg|_{r=r_0}} \,, \label{Aprforce}
\end{eqnarray}
which can be obtained directly using Eq. (\ref{gausscond}),
\begin{equation}
\langle {\bf r}\cdot\ddot{{\bf r}} \rangle=-{1/2} \,.
\end{equation}
For a water bag initial distribution~(\ref{eq waterbag}) we define the envelope 
radius as $r_e=R\sqrt{2}$, so that for $t=0$ $r_e(0)=1$, and  rewrites~(\ref{envelope1}) as
\begin{equation}
\ddot{r}_e(t)+\frac{1}{r_e(t)}=\frac{\varepsilon^2(t)}{r_e^3(t)}\;, \label{envelope2}
\end{equation}
This is the envelope equation.  If initially, $ \langle v^2 \rangle=1/2$ then $\ddot{r}_e=0$
and the envelope will not oscillate.  This is precisely the virial condition.

\clearpage
\section{Potential for Core-Halo Distribution}
\label{Constants}
Integrating the core-halo distribution function over velocity, the dimensionless Poisson 
equation~(\ref{eq poisson}) becomes
\begin{eqnarray}
\label{eq poisson2}
\nabla^2\psi=4\pi^2\left\{
\begin{array}{l}
\eta(\epsilon_F - \psi)+\chi(\epsilon_R-\epsilon_F)\>\, \text{ for } \psi <  \epsilon_F  \\
\chi(\epsilon_R-\psi)\>\, \text{ for } \epsilon_F \le \psi \le \epsilon_R \\
0\>\, \text{ for } \psi > \epsilon_R \,.
\end{array}
\right.
\end{eqnarray}
We define  $\psi_{core}$ for $\psi <\epsilon_F$, $\psi_{halo}$ 
for $\epsilon_F \le \psi \le \epsilon_R$, and $\psi_{out}$ for $\psi > \epsilon_R$. Changing variables
$r^*=2\pi r\sqrt{\eta}$ and 
$r^{**}=2\pi r\sqrt{\chi}$, we can rewrite~(\ref{eq poisson2}) as
\begin{eqnarray}
\psi_{core}'' + \frac{\psi_{core}'}{r^*}+\psi_{core}=\epsilon_F+\frac{\chi}{\eta}(\epsilon_R-\epsilon_F) 
\nonumber\\
\psi_{halo}'' + \frac{\psi_{halo}'}{r^{**}}+\psi_{halo}=\epsilon_R \nonumber\\
\psi_{out}'' + \frac{\psi_{out}'}{r}=0
\end{eqnarray}
The solution of the first two of these equations can be written 
in terms of the Bessel functions of first type and of order $0$,
\begin{eqnarray}
\psi_{core}(r)&=&\epsilon_F+\frac{\chi}{\eta}(\epsilon_R-\epsilon_F)+C_1\>J_0(r^*)+C_{1'}\>Y_0(r^*)\\
\psi_{halo}(r)&=&\epsilon_R+C_2\>J_0(r^{**})+C_3\>Y_0(r^{**})
\end{eqnarray}
The last equation is solved by
\begin{equation}
\psi_{out}(r)=C_4\>\ln r + C_{4'}\,,
\end{equation}
where $\{C_i\}$ are the integration constants.   
The regularity of solution at the origin and Eq.(\ref{gausscond}) require that $C_{1'}=0$, 
$C_{4'}=0$ and $C_4=1$, respectively. The potential reduces to
\begin{eqnarray}
\psi_{core}(r)&=&\epsilon_R+C_1[(\eta/\chi-1)J_0(r_c^*)+J_0(r^*)]\\
\psi_{halo}(r)&=&\epsilon_R+C_2\>J_0(r^{**})+C_3\>Y_0(r^{**})\\
\psi_{out}(r)&=&\ln \left(r\right) \;,
\end{eqnarray}
where we have defined $r_c$ such that $\epsilon_F=\psi(r_c)$.
The others requirements to continuity of the potential and its derivative are,
\begin{eqnarray*}
\left\{ \begin{array}{l}
\psi_{core}(r_c)-\psi_{halo}(r_c)=0\\
\psi_{halo}(r_R)-\psi_{out}(r_R)=0\\
\psi_{halo}'(r_R)-\psi_{out}'(r_R)=0\\
\end{array}
\right.
\end{eqnarray*}
Solving these equations yields the integration constants $C_{1,2,3}$ 
as a function of  $(r_c,\chi)$ and the parameters $(\epsilon_R,\eta)$,
\begin{eqnarray}
\label{C1}
C_1&=&\frac{\pi  \chi  \left(Y_0\left(2\pi r_c\sqrt{\chi }\right) J_0\left(2 \pi r_R \sqrt{\chi }\right)-
J_0\left(2 \pi r_c \sqrt{\chi }\right) Y_0\left(2 \pi r_R \sqrt{\chi }\right)\right)}{2 \eta 
J_0\left(2 \pi r_c \sqrt{\eta }\right)}\\
C_2&=&-\frac{\pi  Y_0\left(2 \pi r_R \sqrt{\chi }\right)}{2}\\
C_3&=&\frac{\pi  J_0\left(2 \pi r_R \sqrt{\chi }\right)}{2}\;,
\end{eqnarray}
The remaining equation of continuity of $\psi'(r)$ at $r_c$ and the conservation of energy will 
determine $r_c$ and $\chi$, see Eq.(\ref{econ}).
\appendix

\end{document}